\documentclass[prl,floats,aps,showpacs,twocolumn]{revtex4}
\usepackage{graphicx}

\font\eightln=line10 at8pt
\catcode`\@=11
\def\singlearrow{\@ifnextchar [{\@singlearrow }{\@singlearrow[0]}}
\def\@singlearrow[#1]{\mathrel{\,\lower0.15ex
  \hbox{\let\@linefnt\eightln\unitlength0.6ex\begin{picture}(4,3)
           \put(0,1.5){\vector(1,0){4}}
    \end{picture}}\,}}
\catcode`\@=12

\font\eightln=line10 at8pt
\catcode`\@=11
\def\doublearrow{\@ifnextchar [{\@doublearrow }{\@doublearrow[0]}}
\def\@doublearrow[#1]{\mathrel{\,\lower0.15ex
  \hbox{\let\@linefnt\eightln\unitlength0.6ex\begin{picture}(4,3)
  \ifcase#1\put(4,0.8){\vector(-1,0){4}}\put(0,2){\vector(1,0){4}}
  \or      \put(0,0.8){\vector(1,0){4}}\put(0,2){\vector(1,0){4}}
  \or      \put(4,0.4){\vector(-2,1){4}}\put(0,0.4){\vector(2,1){4}}
  \fi
  \end{picture}}\,}}
\catcode`\@=12

\font\eightln=line10 at8pt
\catcode`\@=11
\def\triplearrow{\@ifnextchar [{\@triplearrow }{\@triplearrow[0]}}
\def\@triplearrow[#1]{\mathrel{\,\lower0.15ex
  \hbox{\let\@linefnt\eightln\unitlength0.6ex\begin{picture}(4,3)
  \ifcase#1\put(0,0.3){\vector(1,0){4}}\put(0,1.5){\vector(1,0){4}}
           \put(0,2.7){\vector(1,0){4}}
    \or\put(4,0.3){\vector(-1,0){4}}\put(4,1.5){\vector(-1,0){4}}
           \put(0,2.7){\vector(1,0){4}}
    \or\put(4,2){\vector(-2,-1){4}}\put(4,3){\vector(-2,-1){4}}
           \put(0,3){\vector(4,-3){4}}
    \or\put(4,0){\vector(-1,0){4}}\put(4,3){\vector(-2,-1){4}}
           \put(0,3){\vector(2,-1){4}}

\or\put(0,0){\vector(2,1){4}}\put(4,3){\vector(-2,-1){4}}\put(4,0){\vector(-4,3){4}}
  \fi
  \end{picture}}\,}}
\catcode`\@=12

\font\eightln=line10 at8pt
\catcode`\@=11
\def\tripline{\@ifnextchar [{\@tripline }{\@tripline[0]}}
\def\@tripline[#1]{\mathrel{\,\lower0.15ex
  \hbox{\let\@linefnt\eightln\unitlength0.6ex\begin{picture}(4,3)
  \ifcase#1\put(0,0.3){\line(1,0){4}}\put(0,1.5){\line(1,0){4}}
           \put(0,2.7){\line(1,0){4}}
  \or\put(0,0){\line(2,1){4}}\put(0,1){\line(2,1){4}}
           \put(0,3){\line(4,-3){4}}
  \or\put(0,3){\line(2,-1){4}}\put(0,2){\line(2,-1){4}}
           \put(0,0){\line(4,3){4}}
  \or\put(0,0){\line(4,3){4}}\put(0,1.5){\line(1,0){4.5}}
           \put(0,3){\line(4,-3){4}}
  \or\put(0,0.3){\line(1,0){4}}\put(0,1){\line(2,1){4}}
           \put(0,3){\line(2,-1){4}}
  \or\put(0,0){\line(2,1){4}}\put(0,2){\line(2,-1){4}}
           \put(0,2.7){\line(1,0){4}}
  \fi
 \end{picture}}\,}}
\catcode`\@=12

\begin{document}

\title{Semiclassical Foundation of Universality in Quantum Chaos}

\author{Sebastian M{\"u}ller$^1$, Stefan Heusler$^1$, Petr Braun$^{1,2}$,  
Fritz Haake$^1$, Alexander
Altland$^3$}

\address{$^1$Fachbereich Physik, Universit{\"a}t Duisburg-Essen,
45117 Essen, Germany\\
$^2$Institute of Physics, Saint-Petersburg University, 198504  
Saint-Petersburg, Russia\\
$^3$Institut f{\"u}r Theoretische Physik, Z{\"u}lpicher Str 77, 50937 K{\"o}ln, Germany}

\begin{abstract}

We sketch the semiclassical core of a proof of the so-called
Bohigas-Giannoni-Schmit conjecture: A dynamical system with full
classical chaos has a quantum energy spectrum with universal fluctuations on
the scale of the mean level spacing. We show how in the semiclassical
limit all system specific properties fade away, leaving only ergodicity, hyperbolicity, and combinatorics as agents determining the contributions of pairs of classical periodic orbits to the quantum spectral form factor. The small-time form factor is thus reproduced semiclassically. Bridges between classical orbits and (the non-linear sigma model of) quantum field theory are built by revealing the contributing orbit pairs as topologically equivalent to Feynman diagrams.

\end{abstract}


\date{\today}

\maketitle

Fully chaotic dynamics enjoy ergodicity and thus visit everywhere in the
accessible space with uniform likelihood, over long periods of time.
Even long periodic orbits bring about such uniform coverage. Moreover,
classical ergodicity provides quantum chaos with
universal characteristics.

Given chaos, quantum energy levels are correlated within local few-level
clusters but become statistically independent as their distance grows much larger
than the mean level spacing $\Delta$. The decay of correlations on the scale
$\Delta$ is
empirically found system independent, within universality classes distinguished
by presence or absence of time-reversal (${\cal T}$) invariance \cite{gospel,bible}. The
corresponding universal long-time characteristics act on the Heisenberg
scale $T_H=2\pi\hbar/\Delta$, with $\hbar$ Planck's constant.

Universal spectral fluctuations were conjectured as a
manifestation of quantum chaos two decades ago \cite{BGS}. Now, the semiclassical core
of a proof can be given. Based on Gutzwiller's
periodic-orbit theory \cite{Gutzi},
our progress comes with two surprises: one lies in its
simplicity, the other in the appearance of interesting mathematics
(non-trivial properties
of permutations). Moreover, the often disputed intimate relation between
periodic orbits and quantum field theory is confirmed for good. We thus expect the underlying ideas to radiate beyond spectral fluctuations, like to transport and localization.

Technically speaking, we want to show that each completely hyberbolic classical
dynamics has a quantum energy spectrum with the same fluctuations as  a random-
matrix caricature $H_{\rm RMT}$ of its Hamiltonian,
even though that caricature has nothing in common with the
Hamiltonian but symmetry (absence or presence of ${\cal T}$
invariance). The theory of random matrices (RMT) \cite{gospel,bible,Mehta},
developed  by  Wigner and
Dyson to account for fluctuations in nuclear
spectra yields analytic results for correlators of the
level density $\rho(E)$, by averaging over suitable ensembles of random
matrices. Simplest is the two-point correlator
$\overline{\rho(E)\rho(E')}-\overline{\rho(E)}\;\overline{ \rho(E')}$, where the overlines
denote ensemble average. Its Fourier transform with respect to the
energy difference $E-E'$, called spectral form factor $K(\tau)$, is
predicted by RMT for systems without time reversal invariance (unitary
class) and with that symmetry (orthogonal class) as
\begin{equation}
\label{1}
K_{\rm uni}(\tau)=\tau\;,\quad K_{\rm orth}(\tau)=2\tau-\tau\ln(1+2\tau)\,,
\end{equation}
respectively; here $\tau$ is a time measured in units of the Heisenberg time
$T_H$ ranging in $0\leq\tau\leq1$. Note that the ``orthogonal'' form factor admits
the expansion
$K(\tau)=2\tau-2\tau^2+2\tau^3\ldots$ which converges for
$0\leq\tau\leq\frac{1}{2}$. Leaving larger times for future work we propose to show fidelity of individual chaotic dynamics to (\ref{1}).

Of the many ways of doing RMT averages yielding (\ref{1}),
the quantum field-theoretical non-linear sigma model \cite{sigma,Efe} deserves
special mention since it yields  a $\tau$ expansion of
the form factor equivalent to the semiclassical expansion
to be developed here. The model points to analogies between
hyperbolic dynamics and the motion of electrons in disordered media. In fact, the equivalence of semiclassical and field-theoretic expansions was first suggested in the context of disordered metals \cite{disorder}.

Our starting point is Gutzwiller's representation of the level density of a hyperbolic system by a sum over its classical periodic orbits,
$\rho(E)\propto{\rm Re}\sum_\gamma A_\gamma{\rm e}^{{\rm i} S_\gamma/\hbar}$, with $S_\gamma(E)$
the action and $A_\gamma$ the (dimensionless) stability amplitude of the $\gamma$th
orbit. The form factor $K(\tau)$ is the double sum
\begin{equation}
\label{2}
K_{\rm po}(\tau)=\left\langle\sum_{\gamma,\gamma'}A_\gamma
A_{\gamma'}^*
{\rm e}^{{\rm i}(S_\gamma-S_{\gamma'})/\hbar}\delta\big(\tau-\textstyle{\frac{T_\gamma+
T_{\gamma'}}{2T_H}}
\big)\right\rangle
\end{equation}
with $T_\gamma(E)$ the period of $\gamma$; the angular brackets demand averages over  the energy
and over a time interval $\Delta T\ll T_H$. We aim at evaluating the
periodic-orbit sum (\ref{2}) in the {\it semiclassical limit}
$\hbar\to0,\,T_\gamma\to \infty$ with $T_\gamma/T_H=\rm const$. That limit
and the averages indicated eliminate noise due to orbits with $|S_\gamma-S_{\gamma'}|\gg\hbar$
and purge $K(\tau)$ of system specific features.

For the formal double sum in (\ref{2}) to converge to the RMT prediction (\ref{1}),
it must be structured into contributions from {\it families of orbit pairs},
such that each
term of the $\tau$ expansion of $K(\tau)$ comes from a specific set of families.
The simplest family contains the diagonal
pairs $\{\gamma,\gamma\}$ and, given time reversal invariance,
$\{\gamma,{\cal T}{\gamma}\}$, where ${\cal T}\gamma$ is the time reverse of $\gamma$; it yields
Berry's \cite{Berry} ``diagonal approximation''
$K^{(1)}_{\rm po}=\kappa\tau$ where $\kappa=1$ without and $\kappa=2$ with $\cal T$
invariance, due to the doubling of pairs in the latter case. It is here, when summing over the
``diagonal pairs'' that we first meet the ergodicity of long periodic orbits, through
Hannay's and Ozorio de Almeida's (HOdA) \cite{HOdA} sum rule
$\langle\sum_\gamma|A_\gamma|^2
\delta(\tau-\frac{T_\gamma}{T_H})\rangle=\tau$. In a paradigmatic breakthrough, Sieber and Richter \cite{SR} gave the
family responsible for the $\tau^2$ term of $\cal T$ invariant dynamics; it is on
the basis of their insight that we could find and account for all other families.

We first turn to the {\it unitary class} and propose to demonstrate that all families of orbit pairs individually contributing to higher orders $\tau^n$ collectively cancel for $n>1$. To ease our task we assume two freedoms.

 \begin{figure}[b]
   \centering
   \vspace{-0.2cm}

   \resizebox{3in}{!}{\includegraphics{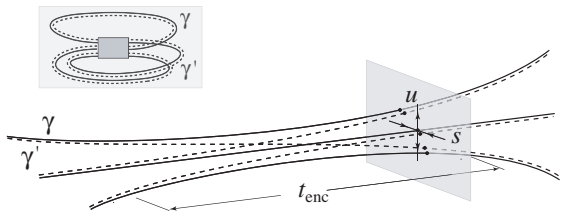}}
   \caption{A triple encounter ($l=3$) in the energy shell and its
   Poincar{\'e} section for an orbit pair $\gamma,\gamma'$. Inset: Global appearance
   of the pair and the generating encounter.}
 \end{figure}

Long orbits have lots of close self-encounters. We speak of
an $l$-encounter when $l$ orbit stretches get and stay close for as long
as their chaotic, i.e. exponential divergence  permits (see Fig.~1). Since the
closest approaches discernible quantum mechanically have an action scale
$\hbar$ we expect relevant encounter durations $t_{\rm enc}$ of the order of the Ehrenfest
time $T_E\sim
\lambda^{-1}\ln\frac{1}{\hbar}$, with $\lambda$ the Lyapounov rate of divergence.
Departing from and ending on the $2l$ ``ports''  of an
$l$-encounter are $l$ ``loops'' with durations of the  order of the
period $T$ and thus of the Heisenberg time $T_H=2\pi\hbar/\Delta=\Omega/2\pi\hbar$, where
$\Omega$ is the volume of the energy shell. Different
encounters must be considered as separate: overlap of any two
would yield a single one with more internal stretches. More generally, an orbit must leave
an encounter before reentering it or another one after traversing an outside loop.

Self-encounters lead us from an orbit $\gamma$
to partners~$\gamma'$. Two orbits in a pair
$\{\gamma,\gamma'\}$ are practically indistinguishable in the loops outside
encounters; they only differ within comparatively short encounters, by their
connections of the outside loops. The action difference~$S_\gamma-S_{\gamma'}$ can thus be of order $\hbar$.
Reshuffling intraencounter connections of $\gamma$ either yields a
partner orbit $\gamma'$ or a pseudo-orbit decomposing into shorter orbits; pseudo-orbits, Fig.~2,
are not admitted to the Gutzwiller sum (\ref{2}).

 \begin{figure}[b]
   \centering
   \vspace{-0.2cm}

   \resizebox{1.5in}{!}{\includegraphics{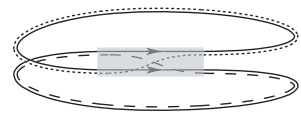}}
   \caption{Illustration of the counting problem: pseudo-orbits (here,
   dashed) must be eradicated.}
 \end{figure}

Calling $v_l$ the number of $l$-encounters ($l\geq2$) within which $\gamma$ and
$\gamma'$ differ by  connections of the coinciding outside loops we write
$V=\sum_lv_l$ for the total number of such encounters
and  $L=\sum_llv_l$ for the number of orbit stretches within encounters
(equalling the
number of loops outside).  We shall see that the
families of orbit pairs with fixed $n=L-V+1$ exclusively contribute to
$\tau^n$ in $K(\tau)$.
To calculate those contributions and check that they sum up to zero for $n>1$
in the absence of $\cal T$ invariance, first
we must understand the
phase-space structure of self-encounters and, second, the
combinatorics of counting proper partners must be mastered.

We begin with a closer analysis of self-encounters \cite{cube}. Drawing a
Poincar{\'e} section, two dimensional for two-freedom
systems, through an $l$-encounter we see the $l$ orbit stretches
of, say, $\gamma$ pierce through that section in $l$ points
$x_i=(u_i,s_i),\,i=1\ldots l$;
one of these can be chosen as the origin of coordinate axes spanned by the
unstable and stable manifolds of $\gamma$ through $x_1=(0,0)$. For an encounter to be
close we require $|u_i|,|s_i|\leq c$, with the bound $c$ small enough for the motion along
the $l$ orbit stretches to allow for mutually linearized treatment.
The $l-1$ piercings $x_i\neq0$ uniquely determine (i) the duration
$t_{\rm enc}^{(l)}(u,s)$ of an $l$-encounter as a (logarithmic) function of the
$u_i,s_i$ \cite{cube}, (ii) the piercings of the
partner orbit(s), and (iii) the contribution to the action difference
$S_\gamma-S_{\gamma'}$. There is a canonical transformation $u,s\to \tilde{u},\tilde{s}$
diagonalizing the action difference to
$\Delta S_{\rm enc}^{(l)}=\sum_{i=2}^l\tilde{u}_i\tilde{s}_i$. Both $t_{\rm  enc}^{(l)}$ and
$\Delta S_{\rm enc}^{(l)}$ are canonical invariants.

We may characterize a set of encounters by a vector $\vec{v}$ whose components
are the numbers $v_l$ of $l$-encounters.
We define a (weighted) number $w(u,s)d^{L-V}\!ud^{L-V}\!s$ of encounter sets with fixed vector
$\vec{v}$ and temporal order of the $L$ visits of the $V$ encounters inside an orbit of
period $T$; it contains a factor for each encounter involved, the fraction of its duration
which the corresponding unstable and stable components spend in the intervals $[u_i,u_i+du_i]$
and $[s_i,s_i+ds_i]$. That number is determined by ergodicity as follows.
A piercing of $\gamma$ through a section will be found with the uniform probability
 $dt_iduds/\Omega$ in a time interval $[t_i,t_i+dt_i]$ and  in the area element
$[u,u+du]\times[s,s+ds]$. We integrate the product of $L-V$ such probabilities
over the $L$ times $t_i$, here $t_1\in [0,T]$ while we restrict the $L-1$
other $t_i$ to (i) a specific order in the interval $[t_1,t_1+T]$ and (ii) by minimal
separations (due to the
ban of encounter overlap). To get the dimensionless weight
$w(u,s)d^{L-V}\!ud^{L-V}\!s$ we  divide the $L$-fold integral by the product
 $\prod t_{\rm enc}^{(l)}$ of durations of the $V$ encounters \cite{cube},
\begin{equation}
\label{4}
w(u,s)=\frac{T\Big(T-\sum\! lt_{\rm enc}^{(l)}\Big)^{L-1}}
{\Omega^{L-V}(L-1)!\prod t_{\rm enc}^{(l)}}\,;
\end{equation}
here, the restrictions mentioned in effect reduce the interval accessible
to $L-1$ integration variables by the cumulative duration $t_{\rm excl}=\sum
lt_{\rm enc}^{(l)}$ of the $L$ intra-encounter stretches. We note that the contribution
$l t_{\rm enc}^{(l)}$ of each $l$-encounter to $t_{\rm excl}$ depends on the
$l-1$ points of piercing $x=(u,s)\neq0$ of $\gamma$ through the pertinent section. We shall
see below that the non-vanishing duration $t_{\rm excl}$ is, even
though a small correction to the period $T$ in $w(u,s)$, of decisive importance for
spectral universality.

Remarkably, $w(u,s)$ results as independent of the order of
the $L$ passages of $\gamma$ through the $V$ self-encounters.
We can therefore proceed to the number of self-encounters
with fixed $\vec{v}$  and $u,s$ irrespective of the order of visits,
$N(\vec{v})w(u,s)$, by accounting for a multiplicity $N(\vec{v})$.

The number $N(\vec{v})$ brings up a combinatorial problem
with a shade of topology mixed in (partner orbits must be connected), decoupled from the phase-space
considerations yielding $w(u,s)$. When stating that
$\gamma$ and $\gamma'$  differ in $v_l$ $l$-encounters, $l=2,3,\ldots$ we
leave open (i) the order in
which the $L$ encounter stretches are passed (in particular, which is the first) and
(ii) how the  intraencounter connections of $\gamma$
are reshuffled in $\gamma'$. The number of possibilities left is
$N(\vec{v})$; we can determine it by running through all different orderings of visits as well
as through all intraencounter connections other than the one realized by $\gamma$
and checking, in each case, whether an orbit or a pseudo-orbit results. For only a few $v_l$
non-zero, $N(\vec{v})$ is easily found with paper, pencil and patience. For general $\vec{v}$,
the permutation problem at issue can be attacked recursively as discussed in the technical
note below.

With the help of the weight $N(\vec{v})w(u,s)$ we replace the sum over
orbit pairs in (\ref{2}) as
$\sum_{\gamma\gamma'}\to \sum_\gamma\!\sum_{\vec{v}}N(\vec{v})\!\!\int\!\!
d^{L-V}\!ud^{L-V}\!s
w(u,s)L^{-1}$. The number of loops $L$ had to be divided out here,
since the $L$ choices of one intra-encounter stretch as the first yield the
same partner orbit. The summand simplifies
as $A_\gamma A_{\gamma'}^*\!\to\!|A_\gamma|^2,\,T_\gamma\!+\!T_{\gamma'}\to
2T_\gamma$ since in contrast
to the action difference $\Delta S$ the prefactors and periods suffer no
relative discrimination by a small quantum unit. We may also invoke the HOdA
sum rule already
met above, to do the sum over $\gamma$, and thus
find the
form factor as
\begin{equation}
\label{5}
\frac{K(\tau)\!-\!\tau}{\tau}=\!\sum_{\vec{v}}N(\vec{v})\!\!\int_{-c}^c\!\!\!\!\!
d^{L-V}\!ud^{L-V}\!s
        \frac{ w(u,s)\,{\rm e}^{{\rm i}\Delta S/\hbar}}{L}.
\end{equation}

Only a single term of the multinomial expansion of $(T-t_{\rm
excl})^{L-1}$ in $w(u,s)$ survives the limit $\hbar\to 0$, the one
which cancels the denominator $\prod t_{\rm enc}^{(l)}$ and
comes with the factor $(T/\Omega)^{L-V}$;
all other terms vanish, either because they involve too low orders in the period $T$
and thus extra factors $\hbar$ besides
a power of $T/T_H$ or because they oscillate rapidly and are annulled by
averaging over a small time window.
The remaining integral becomes elementary after the canonical transformation
diagonalizing $\Delta S$,
$\int_{-c}^c d^{L-V}\!ud^{L-V}\!s\,{\rm e}^{i\Delta S/\hbar}=
(\hbar\int_{-c}^c duds\,{\rm e}^{ i
us/\hbar})^{L-V}\!\to\!(2\pi\hbar)^{L-V}$.
Writing $(T2\pi\hbar/\Omega)^{L-V}=\tau^{L-V}$
we get the series
$K(\tau)=\tau+\sum_{n=2}^\infty K_n\tau^n$ with the coefficient
\begin{equation}
\label{6}
K_n=\frac{1}{(n-2)!}\sum_{\vec{v}}^{(n=L-V+1)}N(\vec{v})\,\frac{(-1)^V\prod_ll^{v_l}}{L}\,.
\end{equation}
governed by ergodicity, combinatorics and topology,
given the existence of sets of separated close self-encounters.

The vanishing of the foregoing sum over families of partner orbits is a
property of the permutation group which to the best of our knowledge was
never noticed before. We sketch
the surprisingly simple proof of $K_n=0$ for $n>0$, based on a recursion scheme for
$N(\vec{v})$, in the technical
note below.  Universal spectral fluctuations are thus established for
dynamics without $\cal T$
invariance.

The {\it orthogonal class} of $\cal T$ invariant dynamics can be treated similarly.
We must generalize the notion of self-encounters to include orbit stretches
close to the
others {\it up to time reversal}. Configuration-space pictures prove useful:
an orbit
stretch may be depicted by an arrow $\singlearrow[0]$. While all
$l$-encounters admitted in the unitary case involve $l$ parallel such arrows (like
$\doublearrow[1]$ or $\triplearrow$) we now face, in addition, arrows with
opposite
directions (like $\doublearrow$ or $\triplearrow[1]$). Likewise, loops in
between self-encounters
of an orbit $\gamma$ appear nearly unchanged in partner(s) $\gamma'$, except
that the senses
of traversal may be opposite.

The multiplicity $N(\vec{v})$ of encounter ``classes'' $\vec{v}$ leads to a
permutation problem
slightly more complicated than in the unitary case. Again, all classes with
fixed $n=L-V+1$
contribute to $\tau^n$. The results (\ref{5}) and (\ref{6}) reappear with an
additional factor 2 due to the
fact that with $\gamma'$ a partner so is ${\cal T}\gamma'$. As discussed in
the technical note, a
recursion relation arises for $K_n$ which yields the random-matrix form
factor for the orthogonal universality class,
$K_{\rm orth}=2\tau+\sum_{n\geq2}\frac{(-2)^{n-1}}{n-1}\tau^n$. We would like to
underscore that in establishing both the unitary and the orthogonal form factor
as universal we have accounted for {\it all} orbit pairs whose members differ by nothing but the way
almost coinciding (up to time reversal) loops are connected within close self-encounters.

The $\tau$ expansion of $K_{\rm orth}$ converges for  $0\leq\tau\leq\frac{1}{2}$. The summed up
logarithm remains valid, by analytic continuation, up to the next singularity. Neither the locus of that singularity ($\tau=1$) nor the form factor for $\tau>1$ can be found within the $\tau$ expansion.
We underscore once more ergodicity and hyperbolicity as our basic assumptions;
in addition, strong action degeneracies as for dynamics with Hecke symmetries must be excluded \cite{Bog}.

We must discuss the relation of our semiclassical work to the
so-called zero dimensional sigma model of quantum
field theory \cite{sigma,Efe,bible}.
The relevance of the sigma model for us lies in
similarities of its perturbative implementation to our semiclassical
expansion. A perturbative evaluation of the sigma model
involves Wick's theorem which
can be shown to entail a recursive reduction scheme equivalent to the
topological and combinatorial
problem yielding our multiplicity $N(\vec{v})$.
Moreover, our orbit pairs correspond to the Feynman
diagrams depicting terms of the
perturbative treatment of the sigma model, with our $l$-encounters and the
outside loops the analogs of
vertices (with $2l$ ports) and propagator lines, respectively. Order by
order in $\tau$, our families of orbit pairs are equivalent to the Feynman diagrams of
the sigma model \cite{disorder}.

{\it Technical note:}
We want to set up the permutation problem yielding the mutiplicity $N(\vec{v})$,
first for the {\it  unitary case}.
To that end, starting from an arbitrary orbit stretch in some encounter we
number the $L$ stretches in the order of visits by $\gamma$. More precisely, we denote
entrance ports of encounters by
$1,2,\ldots, L$ and exit ports by $1',2',\ldots, L'$, such that the $k$-th
stretch of $\gamma$ connects ports $k$ and $k'$. In a partner orbit $\gamma'$ of
 $\gamma$ port $k$ is connected to a port
$j_k'\neq k'$. The $L$ intra-encounter stretches of $\gamma$ thus correspond
to the trivial permutation
$P_E^\gamma=\left({1\atop 1'}{2\atop 2'}{\ldots\atop \ldots}{L\atop
L'}\right)$ while a partner $\gamma'$
will have intra-encounter connections according to $P_E={1,\ldots L\choose j_1',\ldots j_L'}
\neq P_E^\gamma$. Since reconnections take place only within encounters,
$P_E$ must be composed of $v_l$ cycles of length $l, \quad
l=2,3\ldots$. (Mathematically speaking, $P_E$ must belong to the conjugacy class  $2^{v_2}3^{v_3}\ldots l^{v_l}$ of the group of permutations of $L$
objects corresponding to the cycles defined by the vector $\vec{v}$.)

The loops common to $\gamma$ and its partners $\gamma'$ are associated with
the permutation $P_L=\left({1'\atop 2}{2'\atop 3}{\ldots\atop \ldots}{L'\atop 1}\right)$.
The whole of $\gamma$ is represented by the product
$P_E^\gamma P_L=\left({1\atop 2}{2\atop 3}{\ldots\atop \ldots}{L\atop
1}\right)$ and that product
is a single-cycle permutation since $\gamma$ is a single orbit, rather than
a decomposing  pseudo-orbit.
Moreover, the product $P_EP_L$ describes a connected partner
$\gamma'$ rather than a decomposing pseudo-orbit if and only if it is
single-cycle as well. The multiplicity $N(\vec{v})$
is thus found by running $P_E$ through all
possibilities  and counting only those for which $P_EP_L$ is single-cycle.

In  the {\it orthogonal case} there are $2^{l-1}$ distinct orientations of
the stretches of an $l$-encounter. After combining oriented loops with reconnections of
encounter stretches we must again check connectivity.

We have established a general recursion relation for the multiplicity
$N(\vec{v})$, both in the
unitary and the orthogonal case, by following the change of $N(\vec{v})$ as
(i) two encounters unite
($v_l\to v_l-1,v_{l'}\to v_{l'}-1,v_{l+l'-1}\to v_{l+l'-1}+1$, in short
$\vec{v}\to\vec{v}^{[l,l']}$), (ii) as an encounter splits into two
and (iii) as an $l$-encounter becomes an
$(l-1)$-encounter by uniting two of
its orbit stretches.  The relations are  best
written for $\tilde{N}(\vec{v})\equiv N(\vec{v})(-1)^V \prod_l l^{v_l}
[L(n-2)!]^{-1}$; note that
$K_{n}=\sum_{\vec{v}}^{( L-V=n-1)}\tilde{N}(\vec{v})$.

In the unitary case we only need the special variant of the general recursion relation
concerning 2-encounters merging  with $l$-encounters, where the recursion reads
$v_2\tilde{N}(\vec{v})+\sum_{l\geq2}l(v_{l+1}+1)\tilde{N}(\vec{v}^{[l,2]})=0$.
Summing over $\vec{v}$ with fixed $n=L-V+1$ we obtain
$\sum_{\vec{v}}^{(n=L-V+1)}\big[v_2\tilde{N}(\vec{v})+\sum_{l\geq2}l(v_{l+1}+1)
\tilde{N}(\vec{v}^{[l,2]})\big]=0$. In the foregoing double sum we can replace
$\vec{v}^{[l,2]}\to\vec{v}$ and thus $v_{l+1}+1\to v_{l+1}$. The sum now
runs over all $\vec{v}$ with $v_{l+1}>0$; however, the latter restriction is
immaterial due to the factor $v_{l+1}$ in the summand. We thus obtain
$\sum_{\vec{v}}^{(n=L-V+1)}\big[v_2+\sum_{l\geq2}lv_{l+1}
\big]\tilde{N}(\vec{v})=0$; here, the term in the square bracket equals
$n-1$. The resulting identity $(n-1)K_n=0$ implies $K_n=0$ for $n>1$.

In the orthogonal case we need two special cases of the general recursion
relation, to account separately for ``disappearance'' of 2-encounters {\it and} 3-encounters. A
suitable linear combination yields  $K_{n+1}=-\frac{2(n-1)}{n}K_n$; the latter recursion
gives the random-matrix form factor.

Financial support by the SFB/TR12 of the Deutsche Forschungsgemeinschaft is
gratefully acknowledged. We have enjoyed discussions with  J{\"u}rgen
M{\"u}ller, Dominique Spehner
and Martin Zirnbauer.

\end{document}